\documentclass[12pt,a4paper,final]{article}

\usepackage{graphicx}
\usepackage{fancyhdr}

\renewcommand{\baselinestretch}{1.66}

\begin{document}
%TITLE PAGE WITH ABSTRACT
{\renewcommand{\baselinestretch}{1.24}
\title{{\bf Nonlinear quantum optical computing via measurement.}}
\author{{ G. D. Hutchinson$^*$ }\\
{\it Microelectronics Research Centre, Cavendish Laboratory,}\\
{\it  University of Cambridge, Cambridge CB3 0HE, UK.}\\
{ G. J. Milburn}\\
{\it Department of Applied Mathematics and Theoretical Physics,}\\
{\it University of Cambridge, Cambridge UK.} \\
{\it Centre for Quantum Computer Technology,}\\
{\it Department of Physics, The University of Queensland,}\\
{\it St.~Lucia, Queensland 4072, Australia.}\\
{\small $^*$gdh24@cam.ac.uk}\\
{\small Tel: +44 1223 337493 Fax: +44 1223 337706}}
{\date{{\small(Dated: \today})}} \maketitle}

\begin{abstract}
We show how the measurement induced model of quantum computation
proposed by Raussendorf and Briegel [Phys. Rev. Letts. {\bf 86},
5188 (2001)] can be adapted to a nonlinear optical interaction.
This optical implementation  requires a Kerr nonlinearity, a
single photon source, a single photon detector and fast feed
forward. Although nondeterministic optical quantum information
proposals such as that suggested by KLM [Nature {\bf 409}, 46
(2001)] do not require a Kerr nonlinearity they do require complex
reconfigurable optical networks. The proposal in this paper has
the benefit of a single static optical layout with fixed device
parameters, where the algorithm is defined by the final
measurement procedure.
\end{abstract}

\section*{Introduction}
Raussendorf and Briegel \cite{Briegel:01} have shown how efficient
quantum computation can be implemented on an array of suitably
entangled qubits using  a sequence of measurements and conditional
unitary operations conditioned on the measurement results. Their
scheme was based on an array of interacting  spins described by
the Ising Hamiltonian and is thus applicable to certain solid
state implementations. In this paper we show that a similar scheme
can be used in a quantum optical computer based on a mutual Kerr
interaction between optical field modes excited with single photon
states. This interaction was long ago suggested as a possible path
to implementing quantum gates in optics
\cite{Yamamoto88,Milburn89}. While the mutual Kerr interaction is
likely to become a central component in all-optical communication
schemes \cite{Kerr1,Kerr2,Kerr3}, the currently available
technology cannot produce mutual Kerr interactions sufficiently
large to enable single photon operation. Given sufficient
motivation however one might expect that this technological
challenge could be answered and we consider it worthwhile to
investigate simple schemes of all-optical quantum computation that
requires such a nonlinearity. In any case the model we describe
shows that the scheme of Raussendorf and Briegel can be
implemented in principle in an all optical network.

We begin with a brief summary of the scheme in reference
\cite{Briegel:01}. In section 2 we present a dual rail logical
code for qubits based on one photon and two modes per qubit. This
code was recently used by  Knill et al. \cite{Knill:01} to show
that linear efficient quantum computation can be done with linear
optics and single photons provided efficient single photon
measurement and fast feed forward control can be implemented. The
scheme we describe here also uses measurement and feed forward,
but requires a nonlinear interaction. In section 3 we discuss the
practicality of the scheme and end with a short conclusion.

\section{Quantum computing via measurement.} In order to fix
concepts and notation we begin with a brief description of the
Raussendorf and Briegel scheme \cite{Briegel:01}. This scheme
consists of a protocol for teleportation and the implementation of
a controlled not gate (CNOT), where the manipulation of single
qubits via unitary operations are assumed. Consider an array of
qubits that interact via the Ising nearest neighbour interaction,
\begin{equation}
    \mathcal{H} = -\frac{1}{4}g(t)\sigma_{z}^{(1)}\sigma_{z}^{(2)}.
    \label{IsingHam}
\end{equation}
where $\sigma_z^{(i)}$ is the Pauli matrix diagonal in the logic
code (computational)  basis $|0\rangle,|1\rangle$. Here the
strength $g(t)$ can be controlled externally. If the array begins
as a product state in the logic basis, this interaction cannot
create entanglement. If however we initialise the array as a
product state of $\sigma_x^{(i)}$ eigenstates, $|+\rangle =
1/\sqrt{2}(|0\rangle + |1\rangle)$ for each qubit, entanglement
can be created if the Ising interaction is turned on for a certain
time, thus applying the unitary operator
\begin{equation}
    S = \exp ({-i\mathcal{H}_{int}(t)}) .
    \label{StateOperator}
\end{equation}
This initialisation procedure requires only single qubit
rotations, which we assume can be easily implemented. We also note
that for this entangling procedure Raussendorf and Briegel in
their scheme used the Hamiltonian
\begin{equation}
    \mathcal{H}_{\mathrm{int}}=g(t)\sum_{\langle a,a'\rangle}
    \frac{1+ {\sigma_{z}}^{(a)}}{2}\frac{1+ {\sigma_{z}}^{(a')}}{2}.
    \label{TheOtherHam}
\end{equation}
Instead we have used an equivalent but different form
\begin{equation}
    \mathcal{H}_{\mathrm{int}}=-\frac{1}{4}g(t)\sum_{\langle
    a,a'\rangle}{\sigma_{z}}^{(a)}{\sigma_{z}}^{(a')},
    \label{TheOtherHam2}
\end{equation}
in order to implement this entangling operation in our nonlinear
quantum optical computing via measurement scheme.

The time necessary to maximally entangle the qubits can be found
by taking the partial trace over the first qubit, in a two qubit
system. This result generalises to many qubits. If we take the
input state $| \psi\rangle_{in} = |+\rangle_1 \otimes
|+\rangle_2$, apply the unitary operator $U(\theta ) = \exp (
{-i\theta \sigma_{z}^{(1)}\sigma_{z}^{(2)}})$ and take the partial
trace over the second qubit; we have the state
\begin{equation}
    \rho^{(1)}=\phantom{+}|0\rangle\langle 0|+|1\rangle\langle 1| +2\cos
(2\theta)[|1\rangle\langle 0| +|0\rangle\langle 1|].
\end{equation}
Clearly the qubits are maximally entangled when $\theta = \pi / 4$.

To transport an arbitrary state $|\psi_{\mathrm{in}}\rangle =
\alpha |0\rangle + \beta |1\rangle$ across three or more of the
qubits we employ the following protocol:
\begin{enumerate}
    \item We take a chain of an odd number of qubits and prepare them
    in the state $|\psi_{\mathrm{in}}\rangle_{1} \otimes |+\rangle_{2}
    \otimes \ldots \otimes |+\rangle_{n}$ and entangle the qubits
    by applying the unitary operation $ S = \exp \left({-i
    \frac{\pi}{4} \sum_{\langle a,a' \rangle}\sigma_{z}^{(a)} \sigma_{z}^{(a')}}\right)$.

    \item To teleport the state, $\sigma_{x}$ measurements are made
     sequentially on the first qubit through to the second last qubit.
     The process yields the measurement outcomes $s_j \in \{0,1\}$
     for all but the last qubit. These outcomes correspond to the
     qubits being in the states $|+\rangle = |0\rangle_{x}$ and
     $|-\rangle=|1\rangle_{x}$ respectively. The final state of the
     system after the $(n-1)$ $\sigma_{x}$ measurements place the system
     in the state $|s_1\rangle_{1} \otimes |s_2\rangle_{2} \otimes \ldots
     \otimes |s_{n-1}\rangle_{n-1}\otimes |\psi_{\mathrm{out}}\rangle_{n}$.

    \item The output state $|\psi_{\mathrm{out}}\rangle$ is related to the
     input state $|\psi_{\mathrm{in}}\rangle$ by a unitary transformation
     $U_{\mathrm{teleport}} \in \{1,\sigma_{x},\sigma_{z},\sigma_{x}\sigma_{z}
     \}$ determined by the set of measurement outcomes $\{s_{1},s_{2},\ldots,s_{n-1} \}$.
\end{enumerate}
For the case of teleporting over three qubits we find for the
measurement outcomes $\{|-\rangle,|-\rangle\}$,
$\{|+\rangle,|+\rangle\}$, $\{|+\rangle,|-\rangle\}$ and
$\{|+\rangle,|-\rangle\}$ on the first and second qubits require
the unitary operations $1$, $\sigma_{x}$, $\sigma_{z}$ and
$\sigma_{x}\sigma_{z}$ respectively.

\begin{figure}[hbtp]
    \center
    \includegraphics[scale=0.966]{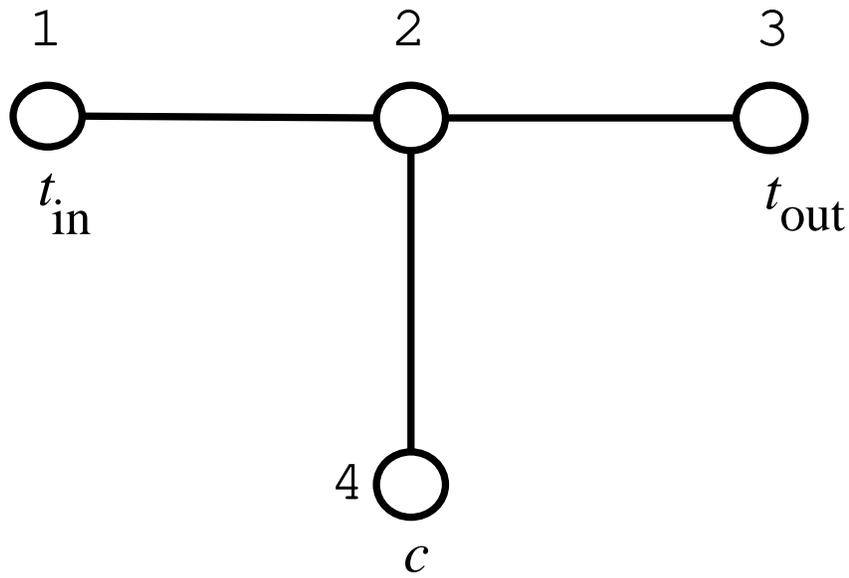}
    \caption{CNOT gate configuration. The CNOT consists of four qubits,
    and uses the measurement of qubits one and two to project an output
    state to qubit three. This state is a known unitary transformation from
    the desired CNOT target state, defined by the two measurement results.
    The CNOT input state $t_{\mathrm{in}}$ is at qubit one and the output
    state $t_{\mathrm{out}}$ is at qubit three. The control state $c$ is
    at qubit four.}
    \label{CNOTconfig}
\end{figure}

By making appropriate measurements on an array of four qubits,
entangled by this scheme, it is possible to implement a particular
universal two qubit gate, the CNOT gate. Consider the two
dimensional array shown in figure \ref{CNOTconfig}. The control
qubit, is labelled $c$ while the target qubits is labelled $t$. In
this scheme the physical system coding the target qubit will
change. We allow for this by the notation $t_{in},t_{out}$. In a
CNOT gate, the logical state of the control does not change.
However if the logical state of the control is $1$ the logical
state of the target qubit is inverted. If the logical state of the
qubit is $0$, the logical state of the target is not changed.

{\bf [Insert figure 1 about here ]}

To implement this gate we place four qubits in the configuration
of figure \ref{CNOTconfig}. Each qubit is a node in this graph
while non-zero nearest neighbour interactions are indicated by
edges. Qubit four is the control. During the operation of the CNOT
in this scheme, qubit $c$ acts as the site where the control state
is input and output at the end of the gate operation. The
information state is input onto qubit $t_{\mathrm{in}}$ and after
the gate operation has completed we have the desired state output
to qubit $t_{\mathrm{out}}$. Our operation of the CNOT gate is as
follows:
\begin{enumerate}
    \item We prepare the states on qubit one and four, these are $| i_{1}\rangle_{1}$ and $| i_{4}\rangle_{4}$
    respectively, where $i_{1}$ and $i_{4}\in \{0,1\}$.
    The other two qubits are prepared in eigenstates of $\sigma_x$ with
    positive eigenvalue, $|+\rangle$. Thus the initial state of the
    CNOT gate is
    \begin{equation}
         |\psi_{\mathrm{in}}\rangle =|i_{1}\rangle_{1} \otimes |+\rangle_{2} \otimes |+\rangle_{3} \otimes |i_{4}\rangle_{4}\ .
    \end{equation}

    \item We entangle the four qubits by turning on the Ising
    interaction between the connected neighbouring sites for an appropriate amount of time $\pi /4$.
    This transforms the initial state of the qubits according to the evolution operator
    \begin{equation}
        S = \exp \left({-i \frac{\pi}{4} \left(\sigma_{z}^{(1)}\sigma_{z}^{(2)} +
        \sigma_{z}^{(2)}\sigma_{z}^{(3)} +
        \sigma_{z}^{(2)}\sigma_{z}^{(4)}\right)}\right).
    \end{equation}
    By entangling the qubits we produced the state
    \begin{eqnarray}
        S|\psi_{\mathrm{in}}\rangle &=&\phantom{+} \exp \left({-i\frac{\pi}{4}((-1)^{i_{1}}+ (-1)^{i_{4}}+1)}\right)|i_{1}00i_{4}\rangle\nonumber\\
        &&+\exp \left({-i\frac{\pi}{4}((-1)^{i_{1}}+(-1)^{i_{4}}-1)}\right)|i_{1}01i_{4}\rangle \nonumber\\
        &&+\exp \left({-i\frac{\pi}{4}((-1)^{1-i_{1}}+ (-1)^{1-i_{4}}-1)}\right)|i_{1}10i_{4}\rangle \nonumber\\
        &&+\exp \left({-i\frac{\pi}{4}((-1)^{1-i_{1}}+ (-1)^{1-i_{4}}+1)}\right)|i_{1}11i_{4}\rangle,
        \label{CNOTstate}
    \end{eqnarray}
    where we have used a more compact notation using the four digits
    in the ket as a representation for the each of the four qubits
    states ordered from qubit 1 to 4.

    \item To output the correct state at qubit three, a $\sigma_{x}$ measurement is made
    on qubit one and then another measurement on qubit two. The measurement outcomes are labelled, $s_j \in \{0,1\}$.

    \item Finally, the output state of the system after the two measurements in the $\sigma_{x}$ basis is
    \begin{equation}
         |\psi_{\mathrm{out}}\rangle =|s_{1}\rangle_{1} \otimes |s_{2}\rangle_{2} \otimes U_{\mathrm{CNOT}}|i_{1}+i_{4}\mathrm{ mod } 2\rangle_{3} \otimes |i_{4}\rangle_{4},
    \end{equation}
    where we have the unitary operator
    \begin{equation}
        U_{\mathrm{CNOT}} = \frac{1}{\sqrt{2}}(1+(-1)^{s_{2}}i)\exp \left({-i\frac{\pi}{4}(1+\sigma_{y}^{(3)})}\right)\left( {\sigma_{x}^{(3)}}\right)^{s_{2}}
        \label{CNOTunitaryTransform}
    \end{equation}
    that acts on the $t_{\mathrm{out}}$ state from the CNOT gate. Up to a local unitary operation we have implemented the
    CNOT gate where the local unitary \ref{CNOTunitaryTransform} is completely determined by the known measurement results.
\end{enumerate}

The form of $U_{\mathrm{CNOT}}$ which acts on qubit three is
\begin{equation}
    U_{\mathrm{CNOT}}
    =\frac{1}{2}(1+(-1)^{s_{2}}i)\exp\left({-i\frac{\pi}{4}}\right)
    \left(
    \begin{array}[c]{cr}
        1 & -1\\
        1 &  1
    \end{array}
    \right) \left(
    \begin{array}[c]{cc}
        0 & 1\\
        1 & 0
    \end{array}
    \right)^{s_{2}},
\end{equation}
and is determined by performing the above scheme. It can be easily
shown why such a unitary operator is necessary. For instance if we
prepare the four qubits of the CNOT configuration in the state
\begin{equation}
    |\psi_{\mathrm{in}}\rangle =|0\rangle_{1} \otimes |+\rangle_{2} \otimes |+\rangle_{3} \otimes |0\rangle_{4},
\end{equation}
then after applying $S$ we have the state
\begin{equation}
    S|\psi_{\mathrm{in}}\rangle = \frac{1}{2} \exp \left({ -i\frac{\pi}{4}}\right) (-i|0000\rangle + i|0010\rangle -|0100\rangle +i|0110\rangle).
\end{equation}
The operation of the CNOT gate should produce the state
$|0\rangle$ on both the control and $t_{\mathrm{out}}$ qubits. By
making $\sigma_{x}$ measurements on qubits one and two we have the
four possible output states shown in table \ref{input00table}. By
following the same procedure for the other three input states we
can determine all the outcomes dependent on all the possible
measurement out comes in tables \ref{input01table} to
\ref{input11table}.\\{\bf [Insert table 1,2,3 and 4 about here ]}

By examination of these results we see that after the two
$\sigma_{x}$ measurements have been made in all sixteen cases the
qubit $c$ is in the desired output state, but qubit
$t_{\mathrm{out}}$ is in a superposition state. There is however a
relationship between the desired output state of the CNOT, the two
$\sigma_{x}$ measurements and the state of the four qubits in the
CNOT configuration. We see that by the application of
$U_{\mathrm{CNOT}}$ to the measurement dependant output on qubit
$t_{\mathrm{out}}$, we recover the desired operation of the CNOT
gate. Finally we note that this relationship is different to the
Raussendorf and Briegel scheme which proposes that the states on
qubits three and four are related to the desired output states via
the operator
\begin{equation}
    U_{\Sigma}^{(34)} = {\sigma_{z}^{(3)}}^{s_{1}+1}{\sigma_{x}^{(3)}}^{s_{2}}{\sigma_{z}^{(4)}}^{s_{1}}.
\end{equation}
This difference arises since we have used a different form of the
Hamiltonian (\ref{TheOtherHam2}) which describes the Ising
interaction. This means that there are the extra single qubit
rotations
$\exp\left(-i\frac{\pi}{4}\right)\left({\sigma_{z}}^{(a)}+
{\sigma_{z}}^{(a')}\right)$ which arise in the evolution operator.

\section{Optical Implementation.}
To adapt the Raussendorf and Briegel scheme to an optical context
we first need to determine the physical states which will carry
the logical code, identify an interaction Hamiltonian that will
provide a similar entangling mechanism, and determine how
appropriate qubit measurements may be made. Our scheme relies on
the dual rail representation of  a single photon excitation of two
distinct  modes. Here we are interested in the presence or absence
of a single photon  in either of the two modes. We represent this
in terms of the number states for each mode labelled a and b,
namely $|0\rangle_{\mathrm{a}}$ and $|1\rangle_{\mathrm{a}}$ for
no photons and a single photon in the a mode respectively. We use
this representation for our qubits, we assign
$|1\rangle_{\mathrm{L}}$ to the presence of the b mode photon.
That is $|1\rangle_{\mathrm{L}} =
|0\rangle_{\mathrm{a}}|1\rangle_{\mathrm{b}}$ and similarly we
have that $|0\rangle_{\mathrm{L}} =
|1\rangle_{\mathrm{a}}|0\rangle_{\mathrm{b}}$. The arbitrary
superposition state $\alpha|0\rangle_{\mathrm{L}}
+\beta|1\rangle_{\mathrm{L}}$ of a qubit in this set up can be
created by making the state $|1\rangle_{\mathrm{L}}$ incident on
an appropriate beam splitter \cite{Knill:01}.

Measurement in the computational basis requires single photon
detection. In the dual rail code, a measurement of $\sigma_z$
requires that the photon detector reliably determine which of the
two modes carries a single photon. As shown in the previous
section however we will need to be able to make measurements in
bases other than the computational basis, in particular we need to
make measurements of $\sigma_{x}$. To achieve this, the qubit
state before the measurement is rotated by $\pi/2$ around the
$\sigma_y$ axis, so that a measurement of $\sigma_z$ after the
rotation is equivalent to a measurement of $\sigma_x$ before the
rotation. As shown by Knill et al. such rotations are easily
implemented by linear optical elements \cite{Knill:01}.

The Hamiltonian which plays the role of the Ising interaction in \cite{Briegel:01} is
\begin{equation}
    \mathcal{H}_{\mathrm{int-opt}} = -(1-2a^{\dagger}a)(1-2b^{\dagger}b),
    \label{OptHam}
\end{equation}
where $a$ and $b$ are the usual annihilation operators for the two
electromagnetic field modes (denoted by a and b say). The terms
$b^{\dagger}b$ and  $a^{\dagger}a$ represent single mode phase
shifts and could be simply photon beam propagation in the a and b
modes. The interaction term represents a mutual, intensity
dependent, phase shift. Such interactions describe a Kerr optical
nonlinearity and arise from a third order nonlinear susceptibility
\cite{quantum-Kerr}. They are routinely used for all optical fibre
switches in optical communication schemes
\cite{Kerr1,Kerr2,Kerr3}. Currently however there is no material
with a sufficiently large Kerr nonlinear susceptibility to
implement a large mutual phase shift at the level of a single
photon, although some progress has been made in this direction
\cite{Kimble} From this Hamiltonian we can define the maximally
entangling evolution operator
\begin{equation}
    \chi = \exp\left({i\frac{\pi}{4} (1-2a^{\dagger}a)(1-2b^{\dagger}b)}\right),\label{EvoluitionChi}
\end{equation}
by analogy to the Hamiltonian used in the Raussendorf and Briegel
scheme. In our optical scheme this is the perfect replacement
since in terms of the number states $\{|0\rangle_{\mathrm{a}}
|0\rangle_{\mathrm{b}}$, $|0\rangle_{\mathrm{a}}
|1\rangle_{\mathrm{b}}$, $|1\rangle_{\mathrm{a}}
|0\rangle_{\mathrm{b}}$, $|1\rangle_{\mathrm{a}}
|1\rangle_{\mathrm{b}} \}$, the action of (\ref{EvoluitionChi}) is
analogous to the action of (\ref{StateOperator}).

\subsection{Optical Teleportation Scheme.}

{\bf [Insert figure 2 about here]}

\begin{figure}[hbtp]
    \center
    \includegraphics[scale=0.640]{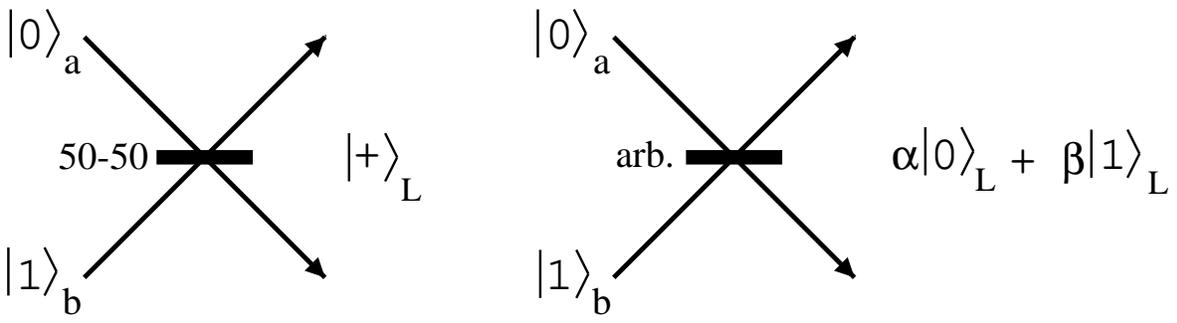}
    \caption{Diagrammatic representation of beam
    splitter output. By the input of a single photon in mode b
    and no photon in mode a, the output from a 50-50 beam splitter
    gives the $|+\rangle_{\mathrm{L}}$ state in the dual rail notation.
    Similarly by inputting single photon in mode b
    and no photon in mode a into an arbitrarily chosen beam splitter
    we have the state $\alpha|0\rangle_{\mathrm{L}}+ \beta|1\rangle_{\mathrm{L}}$}
    \label{BeamSplitterPic}
\end{figure}

\begin{figure}[hbtp]
    \center
    \includegraphics[scale=0.519]{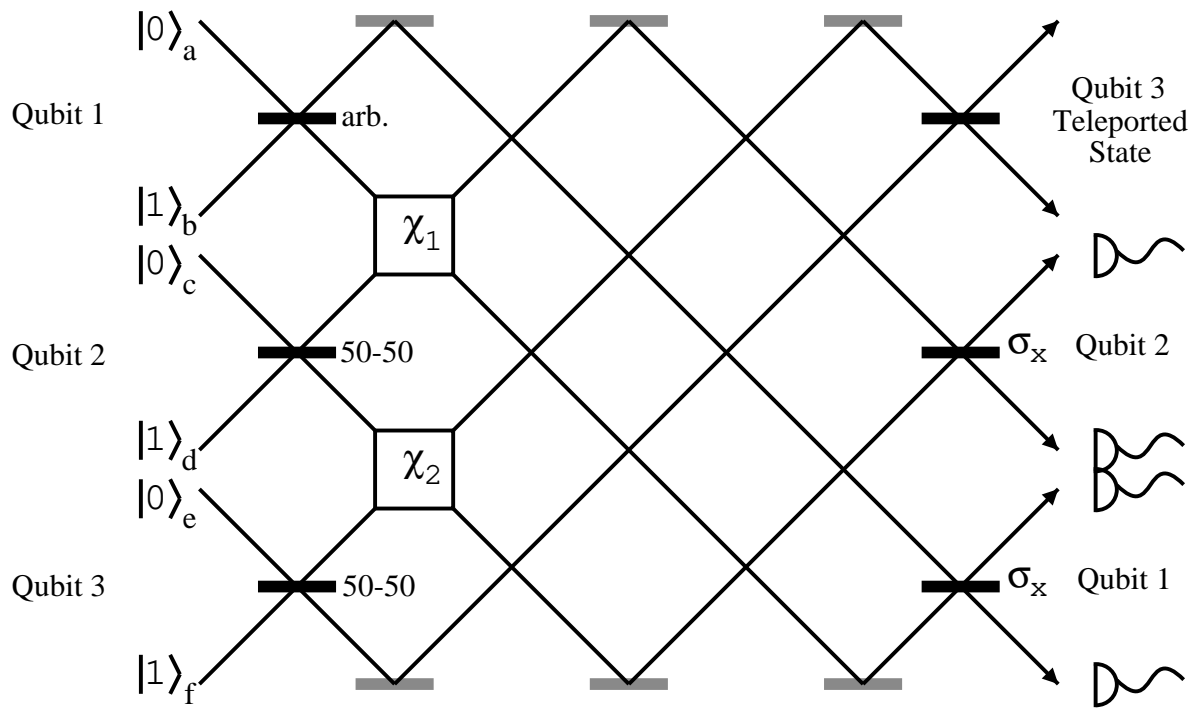}
    \caption{Optical teleportation configuration for three qubits.
    In this scheme we project the state of qubit one to qubit three.
    Qubit one's state can initially be defined by an appropriate
    beam splitter denoted in the figure as an arbitrary choice.
    Qubits two and three are prepared in the state $|+\rangle$
    via a 50-50 beam splitter. The two Kerr nonlinearities $\chi_{1}$
    and $\chi_2$ allow the entanglement of the qubits so that
    the teleportation protocol can be performed. Qubits two and three
    are read out in the $\sigma_x$ basis via single photon detectors
    and appropriate beam splitters, this projects a state onto qubit
    three which is a known unitary transformation from the initial
    input state on qubit one.}
    \label{OptTeleconfig}
\end{figure}

The optical teleportation scheme is shown for three qubits in
figure \ref{OptTeleconfig}. The beam splitters on qubits two and
three are 50-50 beam splitters. This ensures that the state of
qubits two and three is $|+\rangle_{\mathrm{L}}$. At qubit one we
have an appropriate beam splitter that will allow us to create an
arbitrary state such as $\alpha|0\rangle_{\mathrm{L}} + \beta
|1\rangle_{\mathrm{L}}$, this is the state we wish to teleport.
Our diagrammatic representation of the beam splitters is shown in
figure \ref{BeamSplitterPic}.\\{\bf [Insert figure 3 about here]}

To implement the Raussendorf and Briegel scheme we need to
maximally entangle the state
\begin{equation}
    |\psi\rangle_{\mathrm{L}}\otimes |+\rangle_{\mathrm{L}} \otimes |+\rangle_{\mathrm{L}} .
    \label{startstate}
\end{equation}
As we see from figure \ref{OptTeleconfig} we implement this by
putting the input state
\begin{equation}
    |1\rangle_{\mathrm{L}} = |0\rangle_{\mathrm{a}}|1\rangle_{\mathrm{b}}
\end{equation}
into three beam splitters, the first one being chosen to give the
state $|\psi\rangle_{\mathrm{L}}$ and the other two 50-50 beam
splitters. To maximally entangle the three qubits we apply
$\chi_{1}$ to mode a of qubit one and mode d of qubit two. Next we
apply $\chi_{2}$ to mode c of qubit two and mode f of qubit three.
Here $\chi_{1}$ and $\chi_{2}$ are of the form
(\ref{EvoluitionChi}) and we write them as
\begin{eqnarray}
    \chi_{1}&=& \exp\left({i\frac{\pi}{4}(1-2a^{\dagger}a)(1-2d^{\dagger}d)}\right)\\
    \mathrm{ and }\phantom{+}
    \chi_{2}&=& \exp\left({i\frac{\pi}{4}(1-2c^{\dagger}c)(1-2f^{\dagger}f)}\right),
\end{eqnarray}
where the creation and annihilation operators
$a,d,c,f,a^{\dagger},d^{\dagger},c^{\dagger}$ and $f^{\dagger}$
are the operators for the modes denoted by the same subscript a,
d, c and f.

Finally the two beams which constitute each qubit are recombined
for all three of the qubits and by the use of the final two beam
splitters we make $\sigma_{x}$ measurements on qubits one and two
(as shown in figure \ref{OptTeleconfig}). The beam splitter on the
output (qubit three) is used to rotate the final state into the
initial input state $|\psi_{\mathrm{in}}\rangle$ via the feed
forward information provided from the $\sigma_{x}$ measurements.
As discussed in the Raussendorf and Briegel scheme this rotation
is one from the set
$\{1,\sigma_{x},\sigma_{z},\sigma_{x}\sigma_{z}\}$.

By analogy we shall show that this set up gives a one way quantum
computer for photons. By our initial preparation of the three
qubits we have the combined state of the system as
\begin{equation}
    (\alpha|0\rangle_{\mathrm{L}} + \beta|1\rangle_{\mathrm{L}}) \otimes (|0\rangle_{\mathrm{L}} + |1\rangle_{\mathrm{L}})\otimes (|0\rangle_{\mathrm{L}} + |1\rangle_{\mathrm{L}}),
\end{equation}
which we can write in terms of a photon mode representation as
\begin{eqnarray}
    \alpha (|101010\rangle + |101001\rangle +|100110\rangle+ |100101\rangle)\nonumber\\
+\beta (|011010\rangle + |011001\rangle +|010110\rangle+ |010101\rangle).
\end{eqnarray}
Here each of the six entries represent in order the mode photon
number states a, b, c, d, e and f. After the operation of
$\chi_{1}$ we have the state
\begin{eqnarray}
    \alpha (\exp\left({-i\frac{\pi}{4}}\right)|101010\rangle + \exp\left({-i\frac{\pi}{4}}\right)|101001\rangle \nonumber\\
     +\exp\left({+i\frac{\pi}{4}}\right) |100110\rangle+ \exp\left({+i\frac{\pi}{4}}\right) |100101\rangle)\nonumber\\
    +\beta (\exp\left({+i\frac{\pi}{4}}\right) |011010\rangle + \exp\left({+i\frac{\pi}{4}}\right)
    |011001\rangle\nonumber\\
    +\exp\left({-i\frac{\pi}{4}}\right)|010110\rangle +
    \exp\left({-i\frac{\pi}{4}}\right)|010101\rangle),
\end{eqnarray}
by applying $\chi_{2}$ then the state of the configuration is
\begin{eqnarray}
    \alpha (-i|101010\rangle + |101001\rangle +i|100110\rangle+ |100101\rangle)\nonumber\\
+\beta (|011010\rangle +i|011001\rangle +|010110\rangle -i|010101\rangle).
\end{eqnarray}
After taking $\sigma_x$ measurements we can determine the
appropriate rotation to teleport the initial state of the first
qubit onto the third qubit, with the measurement-rotation
relationship corresponding to that stated above. As with the
Raussendorf and Briegel scheme we can readily extend this optical
scheme to any arbitrary number of odd qubits, and similarly to an
even number of qubits.

\subsection{Optical CNOT Scheme.}

{\bf [Insert figure 4 about here]}

\begin{figure}[hbtp]
    \center
    \includegraphics[scale=0.433]{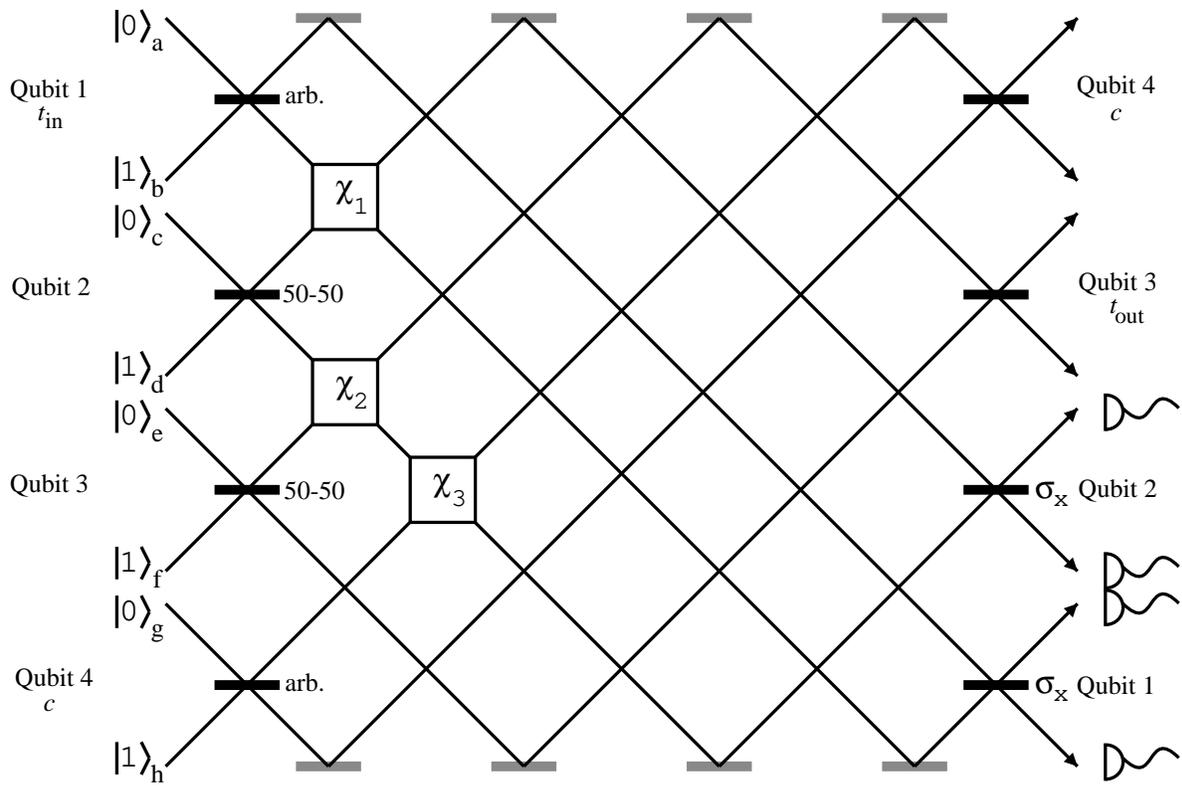}
    \caption{Optical CNOT configuration. Here qubit four and one
    are placed in the desired input states for the CNOT gate
    operation via appropriate beam splitters. Qubits two and
    three are put into the $|+\rangle$ superposition state.
    These qubits are then entangled by the three Kerr nonlinearities
    denoted by $\chi_1$, $\chi_2$ and $\chi_3$. The position of
    these nonlinearities play the role of entangling qubits one,
    three and four to qubit two. Finally $\sigma_x$ measurements
    are made on qubits one and two, projecting the $t_{\mathrm{out}}$
    state to qubit three which is related to desired CNOT output
    by measurement dependant unitary transformations.}
    \label{OptCNOTconfig}
\end{figure}

The optical CNOT scheme is shown in figure \ref{OptCNOTconfig}
where corresponding to the configuration in figure
\ref{CNOTconfig}, we have control qubit $c$ at qubit four, the
$t_{\mathrm{in}}$ at qubit one and $t_{\mathrm{out}}$ at qubit
three. In a similar fashion to the teleportation scheme we
configure the input states using the logical state
$|1\rangle_{\mathrm{L}}$ and the appropriate beam splitter (as
shown in figure \ref{BeamSplitterPic}). By the use of beam
splitters we set qubit one and four to give our desired inputs to
the CNOT. We entangle the four qubits via the consecutive
operation of $\chi_1$, $\chi_2$ and $\chi_3$, and make
$\sigma_{x}$ measurements on qubits one and two via the final beam
splitters. The final output states on qubits three and four can be
rotated via the beam splitters, the actual rotation necessary is
fully specified by the unitary operator
(\ref{CNOTunitaryTransform}) so that the CNOT gate works in the
desired fashion.

For example if we set up the beam splitters on qubits one and four to give the initial state
\begin{equation}
    |\psi_{\mathrm{in}}\rangle_{\mathrm{L}}=|i_{1}\rangle_{\mathrm{L}1} \otimes |+\rangle_{\mathrm{L}2} \otimes |+\rangle_{\mathrm{L}3} \otimes |i_{4}\rangle_{\mathrm{L}4},
\end{equation}
then by the operation of $\chi_{1}$,$\chi_{2}$ and $\chi_{3}$ where
\begin{equation}
    \chi_{3}=\exp\left({i\frac{\pi}{4} (1-2c^{\dagger}c)(1-2h^{\dagger}h)}\right),
\end{equation}
we produce the state
\begin{eqnarray}
    \chi_3 \chi_2 \chi_1 |\psi_{\mathrm{in}}\rangle_{\mathrm{L}} &=& \phantom{+}\exp\left({-i\frac{\pi}{4}((-1)^{i_{1}}+
    (-1)^{i_{4}}+1)}\right)|i_{1}00i_{4}\rangle_{\mathrm{L}}\nonumber\\
    &&+\exp\left({-i\frac{\pi}{4}((-1)^{i_{1}}+ (-1)^{i_{4}}-1)}\right)|i_{1}01i_{4}\rangle_{\mathrm{L}} \nonumber\\
    &&+ \exp\left({-i\frac{\pi}{4}((-1)^{1-i_{1}}+ (-1)^{1-i_{4}}-1)}\right)|i_{1}10i_{4}\rangle_{\mathrm{L}}\nonumber\\
    &&+\exp\left({-i\frac{\pi}{4}((-1)^{1-i_{1}}+(-1)^{1-i_{4}}+1)}\right)|i_{1}11i_{4}\rangle_{\mathrm{L}}.
\end{eqnarray}
This corresponds to (\ref{CNOTstate}), and so justifies our claim
that this scheme is an optical implementation of the Raussendorf
and Briegel Scheme.

\section{Discussion and Conclusion}
The optical implementation of the  scheme  described here requires
a Kerr nonlinearity sufficiently strong so that a single photon
will give a $\pi$ phase shift, single photon sources and very
efficient single photon detection. The latter two requirements,
while difficult, will probably be achieved quite rapidly driven by
a similar need for optical quantum key distribution \cite{Gisin}
and the linear optics quantum computation scheme of Knill et al.
\cite{Knill:01}. Current proposals for single photon sources
include exciton quantum dots
\cite{semiconductor-sources1,semiconductor-sources2,semiconductor-sources3},
NV centres \cite{NV-sources1,NV-sources2}, and surface acoustic
wave devices \cite{Foden}. Current proposals for detectors range
from novel mesoscopic electronic devices \cite{Shields}, and
superconducting devices \cite{superconducting} to novel systems
that use stimulated Raman scattering
\cite{Kwiat-James1,Kwiat-James2} and EIT \cite{munro2003}

The technology that might deliver a Kerr nonlinear device with the
required strength is more difficult to identify. Recent progress
in EIT schemes \cite{EIT1,EIT2,EIT3,EIT4} and cavity QED processes
\cite{Raman1,Raman2} may offer some hope. It might be argued that
if a Kerr nonlinearity with large single photon phase shifts was
available conventional circuit based quantum computing could be
achieved and there would be no need to use the measurement based
scheme in this paper. We wish to emphasis the potential advantages
of the scheme discussed in this paper over conventional gate
network models. A gate network  model based on linear optics and
large Kerr nonlinearities will lead to very complicated
interferometers, with complex choices of the single and two qubit
conditional rotations. On the other hand the scheme described in
this paper requires only a single optical layout with fixed device
parameters. The actual computational algorithm  that one chooses
to perform is left to the measurement scenario. Thus a single
nonlinear optical interferometer implementation  may be used for a
variety of algorithms. This may offer significant technical
advantages and flexibility provided the measurements are very
efficient.

On the other hand, this scheme does require fast, and highly
efficient, photon number measurements, together with an ability to
rapidly apply single qubit unitary operations. While these can be
done with linear optics, the particular linear optics device used
would need to be rapidly controlled by prior photon number
measurements. As photon  detection and control necessarily
requires electronic signal processing, optical delay lines are
likely to be needed. However there is already significant progress
towards photonic quantum memories \cite{Franson-memory} that would
provide a solution here, as for the similar problem in linear
optical quantum computing.

We have already indicated that there is sufficient technical
progress in the area of efficient single photon detection to
believe that the required measurements can be achieved in the near
future. There may be some advantages to trading interferometer
design for complex measurement scenarios provided the required
single photon detection efficiencies are available.

\section*{Acknowledgments} GDH acknowledges the financial support of
the Australian Research Council Special Research Centre for
Quantum Computer Technology. GJM acknowledges the support of the
Cambridge MIT Institute.
\bibliographystyle{plain}
 \clearpage
\begin{table}[hbtp]
    \centering
    \begin{tabular}{@{}cr}
        \hline
        \textbf{Measurement result} & \textbf{Output state}  \\
        \hline
         ($|+\rangle_{1}$, $|+\rangle_{2}$) & $-\frac{1}{2}\exp\left({ -i \frac{\pi}{4}}\right) (1+i)(|+\rangle_{1} \otimes |+\rangle_{2}\otimes (|0\rangle_{3}-|1\rangle_{3})\otimes |0\rangle_{4})$ \\
         ($|+\rangle_{1}$, $|-\rangle_{2}$) & $\frac{1}{2}\exp\left({ -i \frac{\pi}{4}}\right) (1-i)(|+\rangle_{1} \otimes |-\rangle_{2}\otimes (|0\rangle_{3}+|1\rangle_{3})\otimes |0\rangle_{4})$ \\
         ($|-\rangle_{1}$, $|+\rangle_{2}$) & $-\frac{1}{2}\exp\left({ -i \frac{\pi}{4}}\right) (1+i)(|-\rangle_{1} \otimes |+\rangle_{2}\otimes (|0\rangle_{3}-|1\rangle_{3})\otimes |0\rangle_{4})$ \\
         ($|-\rangle_{1}$, $|-\rangle_{2}$) & $\frac{1}{2}\exp\left({ -i \frac{\pi}{4}}\right) (1-i)(|-\rangle_{1} \otimes |-\rangle_{2}\otimes (|0\rangle_{3}+|1\rangle_{3})\otimes |0\rangle_{4})$ \\
        \hline\hline
    \end{tabular}
    \caption{Output states for four possible $\sigma_{x}$ measurements when ($|0\rangle_{1}$,$|0\rangle_{4}$) is input into the CNOT.}
    \label{input00table}
\end{table}
\clearpage
\begin{table}[hbtp]
    \centering
    \begin{tabular}{@{}cr}
        \hline
        \textbf{Measurement result} & \textbf{Output state}  \\
        \hline
         ($|+\rangle_{1}$, $|+\rangle_{2}$) & $\frac{1}{2}\exp\left({ -i \frac{\pi}{4}}\right) (1+i)(|+\rangle_{1} \otimes |+\rangle_{2}\otimes (|0\rangle_{3}+|1\rangle_{3})\otimes |1\rangle_{4})$ \\
         ($|+\rangle_{1}$, $|-\rangle_{2}$) & $\frac{1}{2}\exp\left({ -i \frac{\pi}{4}}\right) (1-i)(|+\rangle_{1} \otimes |-\rangle_{2}\otimes (|0\rangle_{3}-|1\rangle_{3})\otimes |1\rangle_{4})$ \\
         ($|-\rangle_{1}$, $|+\rangle_{2}$) & $\frac{1}{2}\exp\left({ -i \frac{\pi}{4}}\right) (1+i)(|-\rangle_{1} \otimes |+\rangle_{2}\otimes (|0\rangle_{3}+|1\rangle_{3})\otimes |1\rangle_{4})$ \\
         ($|-\rangle_{1}$, $|-\rangle_{2}$) & $\frac{1}{2}\exp\left({ -i \frac{\pi}{4}}\right) (1-i)(|-\rangle_{1} \otimes |-\rangle_{2}\otimes (|0\rangle_{3}-|1\rangle_{3})\otimes |1\rangle_{4})$ \\
        \hline\hline
    \end{tabular}
    \caption{Output states for four possible $\sigma_{x}$ measurements when ($|0\rangle_{1}$,$|1\rangle_{4}$) is input into the CNOT.}
    \label{input01table}
\end{table}
\clearpage
\begin{table}[hbtp]
    \centering
    \begin{tabular}{@{}cr}
        \hline
        \textbf{Measurement result} & \textbf{Output state}  \\
        \hline
         ($|+\rangle_{1}$, $|+\rangle_{2}$) & $\frac{1}{2}\exp\left({ -i \frac{\pi}{4}}\right) (1+i)(|+\rangle_{1} \otimes |+\rangle_{2}\otimes (|0\rangle_{3}+|1\rangle_{3})\otimes |0\rangle_{4})$ \\
         ($|+\rangle_{1}$, $|-\rangle_{2}$) & $\frac{1}{2}\exp\left({ -i \frac{\pi}{4}}\right) (1-i)(|+\rangle_{1} \otimes |-\rangle_{2}\otimes (|0\rangle_{3}-|1\rangle_{3})\otimes |0\rangle_{4})$ \\
         ($|-\rangle_{1}$, $|+\rangle_{2}$) & $-\frac{1}{2}\exp\left({ -i \frac{\pi}{4}}\right) (1+i)(|-\rangle_{1} \otimes |+\rangle_{2}\otimes (|0\rangle_{3}+|1\rangle_{3})\otimes |0\rangle_{4})$ \\
         ($|-\rangle_{1}$, $|-\rangle_{2}$) & $-\frac{1}{2}\exp\left({ -i \frac{\pi}{4}}\right) (1-i)(|-\rangle_{1} \otimes |-\rangle_{2}\otimes (|0\rangle_{3}-|1\rangle_{3})\otimes |0\rangle_{4})$ \\
        \hline\hline
    \end{tabular}
    \caption{Output states for four possible $\sigma_{x}$ measurements when ($|1\rangle_{1}$,$|0\rangle_{4}$) is input into the CNOT.}
    \label{input10table}
\end{table}
\clearpage
\begin{table}[hbtp]
    \centering
    \begin{tabular}{@{}cr}
        \hline
        \textbf{Measurement result} & \textbf{Output state}  \\
        \hline
         ($|+\rangle_{1}$, $|+\rangle_{2}$) & $\frac{1}{2}\exp\left({ -i \frac{\pi}{4}}\right) (1+i)(|+\rangle_{1} \otimes |+\rangle_{2}\otimes (|0\rangle_{3}-|1\rangle_{3})\otimes |1\rangle_{4})$ \\
         ($|+\rangle_{1}$, $|-\rangle_{2}$) & $-\frac{1}{2}\exp\left({ -i \frac{\pi}{4}}\right) (1-i)(|+\rangle_{1} \otimes |-\rangle_{2}\otimes (|0\rangle_{3}+|1\rangle_{3})\otimes |1\rangle_{4})$ \\
         ($|-\rangle_{1}$, $|+\rangle_{2}$) & $-\frac{1}{2}\exp\left({ -i \frac{\pi}{4}}\right) (1+i)(|-\rangle_{1} \otimes |+\rangle_{2}\otimes (|0\rangle_{3}-|1\rangle_{3})\otimes |1\rangle_{4})$ \\
         ($|-\rangle_{1}$, $|-\rangle_{2}$) & $\frac{1}{2}\exp\left({ -i \frac{\pi}{4}}\right) (1-i)(|-\rangle_{1} \otimes |-\rangle_{2}\otimes (|0\rangle_{3}+|1\rangle_{3})\otimes |1\rangle_{4})$ \\
        \hline\hline
    \end{tabular}
    \caption{Output states for four possible $\sigma_{x}$ measurements when ($|1\rangle_{1}$,$|1\rangle_{4}$) is input into the CNOT.}
    \label{input11table}
\end{table}
\clearpage
\clearpage
\clearpage
\clearpage
\end{document}